\documentclass[runningheads]{llncs}
\usepackage{graphicx}
\usepackage{multirow}

\usepackage[table,xcdraw]{xcolor}
\usepackage{cite}
\usepackage{marvosym}
\begin{document}
\title{Pathological Image Segmentation with Noisy Labels}
%
%
%
%
\author{Li Xiao\inst{1}\textsuperscript{\#} \and Yinhao Li\inst{1}\textsuperscript{\#} \and Luxi Qv\inst{2}\textsuperscript{\#} \and Xinxia Tian\inst{3}\textsuperscript{*} \and Yijie Peng\inst{2}\textsuperscript{*} \and S. Kevin Zhou\inst{1}\textsuperscript{*}}
%
\authorrunning{L. Xiao et al.}
\institute{Institute of Computing Technology, Chinese Academy of Sciences\\
\email{xiaoli@ict.ac.cn, liyinhao0413@foxmail.com,s.kevin.zhou@gmail.com}\and
Guanghua School of Management, Peking University\\
\email{qvluxi1997@pku.edu.cn,pengyijie@pku.edu.cn}\and
Peking University Third Hospital\\
\email{tianxinxia@163.com}}

\maketitle              
\begin{abstract}
Segmentation of pathological images is essential for accurate disease diagnosis. The quality of manual labels plays a critical role in segmentation accuracy; yet, in practice, the labels between pathologists could be inconsistent, thus confusing the training process. In this work, we propose a novel label re-weighting framework to account for the \textit{reliability of different experts' labels on each pixel} according to its surrounding features. We further devise a new attention heatmap, which takes \textit{roughness as prior knowledge} to guide the model to focus on important regions. Our approach is evaluated on the public Gleason 2019 datasets. The results show that our approach effectively improves the model’s robustness against noisy labels and outperforms state-of-the-art approaches.

\keywords{Pathology segmentation \and label re-weighting \and noisy labels.}
\end{abstract}
\section{Introduction}

Biopsy refers to a technique of removing diseased tissue from a living body for pathological examination. It is the most reliable way to determine whether an area is cancerous or not. Biopsy generally involves sectioning, H\&E staining, and observing under a microscope. Pathologists examine the regularities of cell shapes, tissue distributions, and other features to determine cancerous regions and the malignancy degree. The diagnostic results of pathological images directly affect surgical decisions, and thus misdiagnosis should be avoided as much as possible. The shortage of pathologists and their heavy workload increase the demand for intelligent diagnosis systems for pathology images. Digitization of whole slide images led to the emergence of artificial intelligence in digital pathology,
which has been reshaping pathology systems\cite{Goldenberg2019A} and improving patient management\cite{0Artificial}. Deep learning frameworks have  achieved remarkable success on many pathology image recognition tasks, such as intelligent diagnosis 
of breast\cite{2016Deep}, lung\cite{Nicolas2018Classification} and prostate cancer\cite{2019Deep}\cite{2020Automated}.

There are many challenges in identifying different grades of cancer in pathological images. Segmenting pathological regions with grading information is critical for automatic auxiliary diagnosis. A critical challenge is that the feature differences between adjacent grades may be tiny.  As a result, accurate annotations for pathology images are difficult to obtain. At the Vienna Consensus Conference, the diagnosis agreement on gastric cancer between Japanese and Western scholars was less than 40\%. It is difficult for both Western and Japanese pathologists to make a repeatable distinction between certain subcategories\cite{R2004Classification}. 
In practice, to reduce inter-observer variability among expert pathologists, labels are  usually determined according to annotations of multiple experts. This poses an interesting question ahead of us: \textit{Is there a way to reduce biases introduced by inter-observer discrepancies, i.e., effective learning from noisy labels}? 

Learning from noisy labels has been a challenge in the field of computer vision with many pioneer models developed. Chen \emph{et al}.~\cite{chen2015webly} separate the datasets into easy and hard images according to the noise level of labeling and introduce a noise adaption layer for classifying hard images. Li \emph{et al}.~\cite{li2019learning} propose a noise-tolerant training algorithm that uses synthetic noisy labels to avoid the model's overfitting for specific noises. 
In \cite{jiang2018mentornet,li2019learning,wang2020noise}, a teacher-student structure is utilized to overcome the negative impact of noisy labels, but they only deal with the situation where one image corresponds to only one label. Yu et al. \cite{yu2020difficulty} use the annotation information of multiple experts and achieve convincing results on the Glaucoma Classification task. In this work, we propose a novel approach to improve the segmentation performances by learning the reliability of different experts' labels on the same pathology image. To the best of our knowledge, this is the first work to learn the variability of multi-expert labels on the segmentation task.

Specifically, we propose a novel label re-weighting framework to account for the \underline{reliability of different experts' labels on each pixel} according to its surrounding features. We then devise a new Gaussian attention focal loss, which takes \underline{roughness as prior knowledge} to represent each pixel's importance, and generate a heatmap to re-weights the focal loss\cite{lin2017focal}. Our model is validated on the public Gleason 2019 datasets\cite{2018Automatic}. The results show that our approach effectively improves the model’s robustness against noisy labels and outperforms state-of-the-art approaches.


\section{Methods}
\subsection{Label Re-weighting Framework}

\begin{figure}[h]
\centering
\includegraphics[scale=0.2]{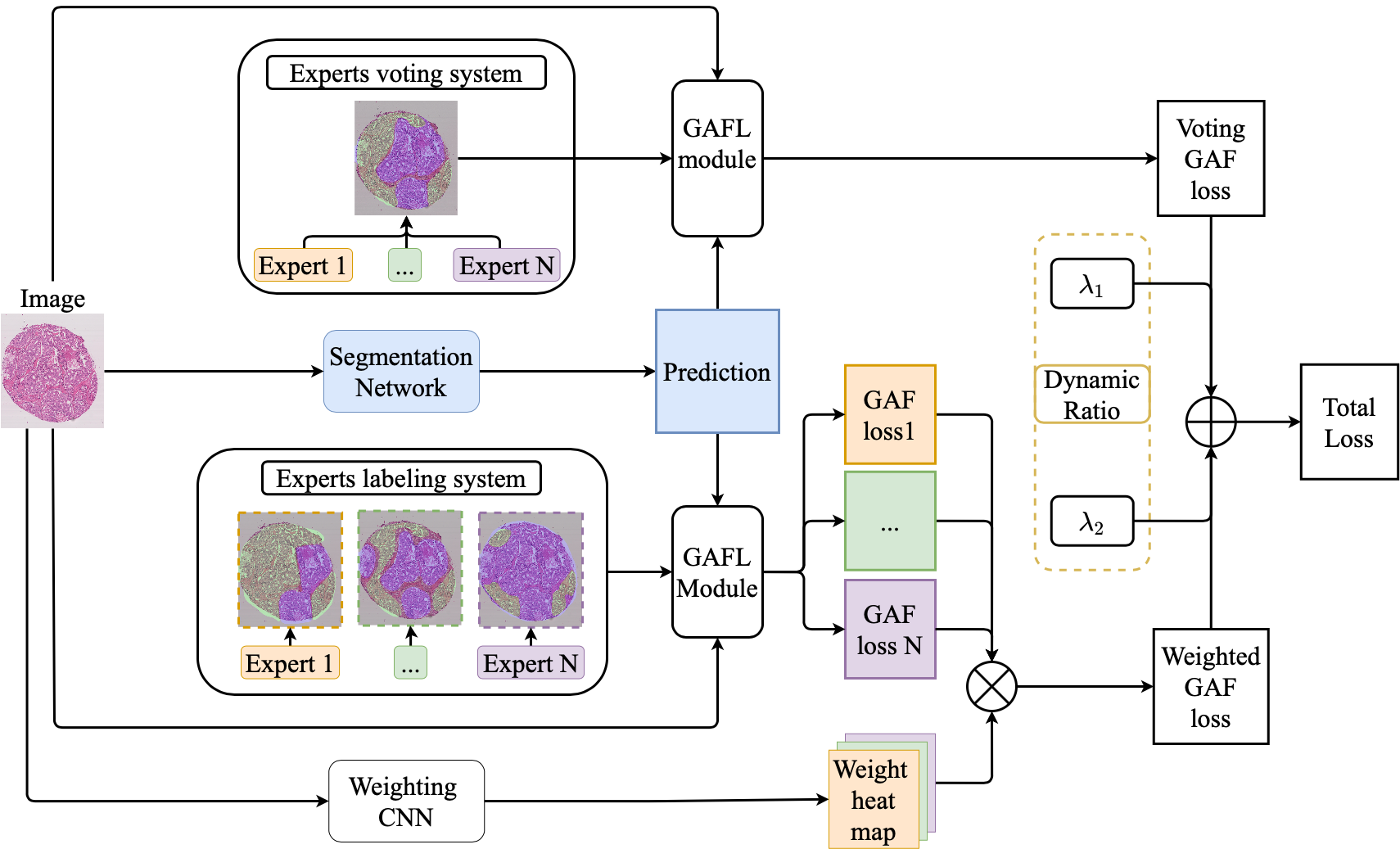}
\caption{Label re-weighting architecture}
\label{label-reweighting}
\end{figure}

Pathologists may have different views on a specified pathological image. Factors such as varying professional levels and personal preference introduce bias from pathologists when labeling the same pathology image. Therefore, we propose a label re-weighting architecture to reduce such errors. Specifically, we introduce a weighting CNN to learn a weight heatmap for assigning weights on each pixel's loss value. The weighting CNN is optimized with respect to the loss, and it learns \underline{the reliability of different experts on each pixel} according to its surrounding features. We further propose a novel Gaussian attention focal loss (GAF loss), which takes \underline{roughness as prior knowledge} and makes the model pay more attention to important regions through a weighting mechanism (as shown in section 2.2 in detail).

The architecture of the label re-weighting framework is shown in Fig.~\ref{label-reweighting}, which consists of a voting loss and a weighted loss. Assume a given image $I$ is annotated by N experts, and let $\{P_1,P_2,\ldots,P_N\}$ be $N$ labels annotated by $N$ experts, and $P$ be the prediction of a network for image $I$. The voted annotation is computed by a major voting strategy according to the $N$ labels:
\begin{equation}
    P_{vote}=major(P_1,P_2,...,P_N)
\end{equation}

 $N$ losses $\{l_1, l_2,...,l_N\}$ are computed directly according to discrepancies between $\{P_1,P_2,...,P_N\}$ and $P$. A voting loss $L_{vote}$ is computed between $P_{vote}$ and $P$.  All the above $(N+1)$ losses are focal losses re-weighted by a Gaussian attention heatmap (denoted as GAFL module, see section 2.2 in detail). A weight heatmap further acts on $\{l_1, l_2,...,l_N\}$ to generate weighted losses. The weighted losses are then concatenated with the voting loss to obtain the final loss. 

The weight heatmap is constructed by a weighting CNN. The weighting CNN $\phi_{weight}$ takes a pathology image $I$ as input and outputs the corresponding weight heatmap: $W=\{w_1,w_2,...,w_N\}$.
\begin{equation}
W_n = \phi_{weight,n}(I),~n=1,2,...,N.
\end{equation}
The weight heatmap has a shape of (N \(\times\) H \(\times\) W), where N stands for N experts, H and W stand for the height and width. We concatenate the N loss and multiply them with the weight heatmap to generate the weighted loss:
\begin{equation}
L_{weighted} = \frac{1}{N} \sum_{n=1}^N l_n \odot w_n.
\end{equation}


The final loss is the weighted sum of weighted loss and voting loss: 
\begin{equation}
    L_{total} =  \lambda_1 L_{vote} + \lambda_2  L_{weighted},
\end{equation}
where $\lambda_1$ and $\lambda_2$ are balancing weights set empirically in practice.

At the beginning of the training process, we set $\lambda_1$ larger than $\lambda_2$ to let the segmentation network converge quickly. As the training progresses, $\lambda_2$ gradually becomes larger, allowing weighting CNN to take effect.

\subsection{Gaussian attention focal loss}
Currently, segmentation loss function, such as cross-entropy, dice loss\cite{milletari2016v}, focal loss\cite{lin2017focal}, treats each pixel equally. There are some ways to improve the performance by selectively optimizing more important regions. For example,  Fan \emph{et al}.\cite{fan2020pranet} improve the performances by adding a weight map generated by annotations.

In pathological diagnosis, regions with different texture information can play different roles in determining cancers and their grades.  As a preliminary exploration using a baseline with the focal loss only for segmentation, most hard samples fall into the rough regions. This means rough regions may play more important roles in determining the grades and need to pay special attention during training. 

\begin{figure}[h]
\centering
\includegraphics[scale=0.2]{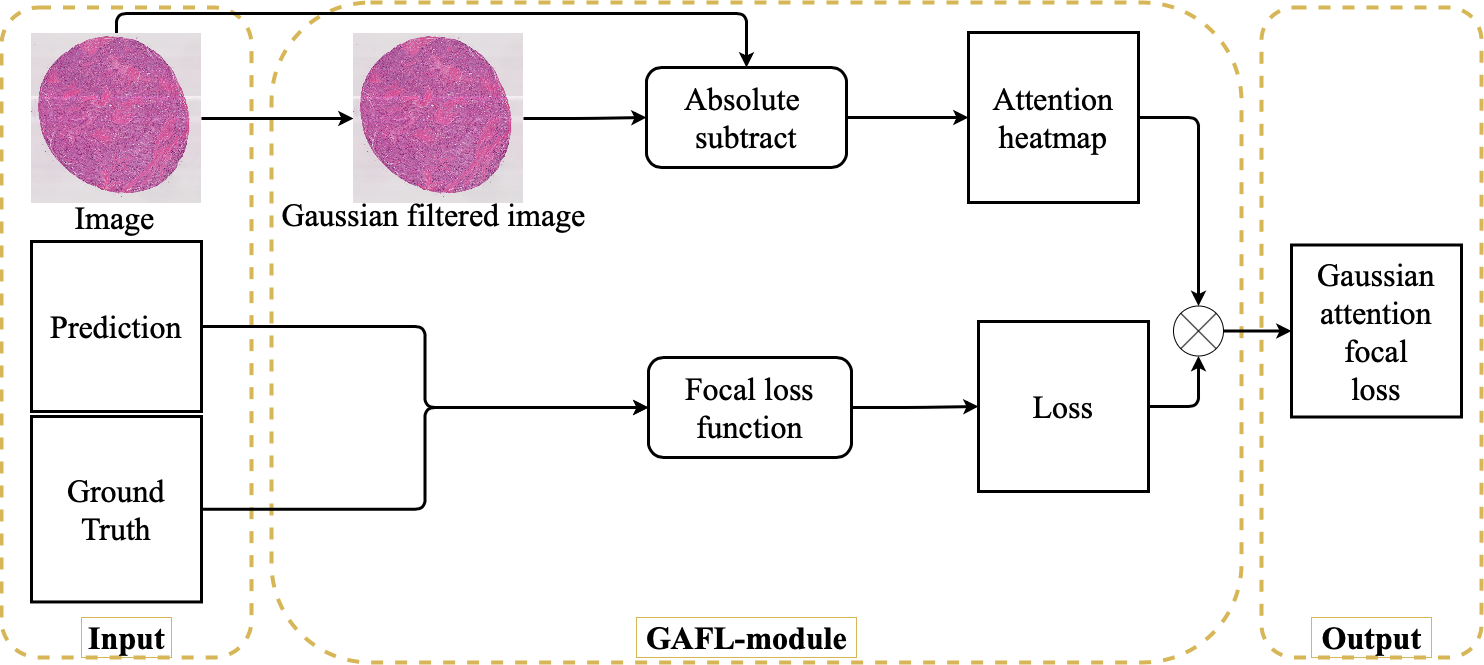}
\caption{GAFL
Module:focal loss with Gaussian attention heatmap}
\label{weight_loss}
\end{figure}

In this work, we propose a Gaussian attention focal loss (GAFL), 
which optimizes 
the focal loss by considering the \underline{roughness of each pixel as prior 
knowledge}. Specifically, as shown in Figure \ref{weight_loss}, for each input $I$, we apply a Gaussian filter $Gauss(I)$, and obtain the attention map $W_{heat}$ according to the absolute value of the difference between the original input and filtered input: 
\begin{equation}
W_{heat}=\lambda_{a}*abs(I-Gauss(I))+\lambda_{b},
\end{equation}
Here $\lambda_a,\lambda_b$ are coefficients to avoid zero pixel values on the heatmap. As a result, the attention map represents the roughness of each pixel and its surroundings. The attention map is further multiplied with the focal loss\cite{lin2017focal} ($L_{focal}=-\sum_{t}{\alpha_{t}(1-p_{t})^{\gamma}p_{t}}$) to obtain the final GAF loss: 
\begin{equation}
L_{GAF} = W_{heat}\odot L_{focal}
\end{equation}

Different than Pranet\cite{fan2020pranet}, the weighted map in our work is directly obtained from the original image rather than annotations and thus can maintain more precise information. The $L_{GAF}$ is used as a GAFL module in the label re-weighting framework (as shown in Figure \ref{label-reweighting}) to obtain the N losses and voting loss.

\section{Experimental Settings}

\subsection{Datasets}
We use the Gleason 2019 dataset \cite{2018Automatic} for the task of segmenting images with noisy labels to validate our model. The dataset is composed of tissue micro-array (TMA) images about Prostate Cancer, including four categories: Benign, Gleason grade 3, Gleason grade 4, and Gleason grade 5. Each TMA image is annotated with numbers 0, 1, 3, 4, 5, 6 in detail by several pathologists. According to the official website, labels 0, 1, and 6 indicate benign (no cancer), and labels 3, 4, and 5 correspond to Gleason grade 3, grade 4 and grade 5, respectively. The number of images is 244. 

\begin{figure}[h]
\centering
\includegraphics[scale=0.5]{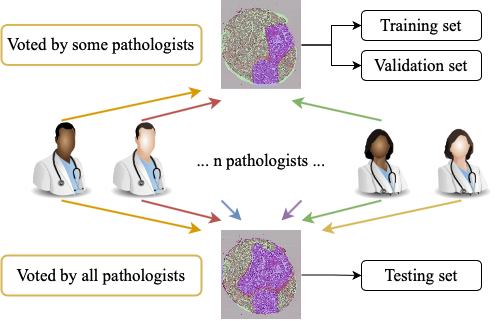}
\caption{Generation process of training,validation and testing set.}
\label{dataset_fig}
\end{figure}

It is worth mentioning that not every image has annotations from all six pathologists.  There are 40 images in the Gleason 2019 dataset that have annotations from all six pathologists. We assume that voting annotations with more pathologists have less bias; thus, we treat these 40 images as the \underline{testing set}. Out of the remaining 204 images, 195 are labeled by the three pathologists numbered 1, 3, and 4, and we choose these 195 images as the training set and the validation set (see Fig. \ref{dataset_fig} for details). 

\begin{table}[h]
\caption{Annotation information of the experimental datasets.}

\begin{tabular}{cccccc}
\hline
 &
   &
  Benign &
  Gleason grade 3 &
  Gleason grade 4 &
  Gleason grade 5 \\ \hline
\multicolumn{1}{c|}{\multirow{4}{*}{\begin{tabular}[c]{@{}c@{}}Training \&\\ validation\\ set\end{tabular}}} &
  \multirow{2}{*}{Ratio of images} &
  \multirow{2}{*}{100.00\%} &
  \multirow{2}{*}{44.62\%} &
  \multirow{2}{*}{78.97\%} &
  \multirow{2}{*}{13.33\%} \\
\multicolumn{1}{c|}{} &
   &
   &
   &
   &
   \\ \cline{2-6} 
\multicolumn{1}{c|}{} &
  \multirow{2}{*}{Ratio of pixels} &
  \multirow{2}{*}{59.63\%} &
  \multirow{2}{*}{13.26\%} &
  \multirow{2}{*}{26.26\%} &
  \multirow{2}{*}{0.85\%} \\
\multicolumn{1}{c|}{} &
   &
   &
   &
   &
   \\ \hline
\multicolumn{1}{c|}{\multirow{2}{*}{\begin{tabular}[c]{@{}c@{}}Testing\\ set\end{tabular}}} &
  Ratio of images &
  100.00\% &
  60.00\% &
  85.00\% &
  15.00\% \\ \cline{2-6} 
\multicolumn{1}{c|}{} &
  Ratio of pixels &
  56.26\% &
  13.41\% &
  28.93\% &
  1.40\% \\ \hline
\end{tabular}
\label{dataset_table}
\end{table}

Table \ref{dataset_table} lists the statistics of our data set. The ratio of images represents the proportion of WSIs that contains a certain category. And the ratio of pixels represents the proportion of the pixels over all images for a certain category. It is worth noting that the dataset is highly unbalanced and it only contains a tiny portion of Gleason grade 5 regions.

\subsection{Implementation Details}
Our model is implemented based on an open-source segmentation structure mmsegmentation \cite{mmseg2020}. All the experiments are implemented using a TITAN-RTX graphics card. Each image is resized to 1024\(\times\)1024, the batch size is set to 2, and the initial learning rate is set to 0.01. The number of training steps is 10k. We add data augmentation operations to the data processing pipeline. The image randomly flips and rotates 90 degrees or 180 degrees or 270 degrees with a probability of 0.5 and randomly resizes within a scale of 0.9-1.1. We develop the weighting CNN using ResNet-18 \cite{he2016deep} as the backbone and replace the fully connected layer with a softmax layer. We use bilinear interpolation to resize the feature map to the same size as the feature map output by the segmentation network. In the Gaussian attention focal loss module, we set the Gaussian kernel with a radius of 5 and a standard deviation of 3, and set $\lambda_a=50, \lambda_b=1$.

During training, we save checkpoints every 1k iterations and verify them on the validation set. When verifying the effectiveness of our loss function, we replace the GAFL module with the corresponding loss function in the framework of Fig.\ref{label-reweighting}. The value of $\lambda_1$ and $\lambda_2$ are also changed as the training process increases. We define a value $n$ to moderate the two parameters. $n$ is initially set to 0, and increases by 1 every 1k iterations:
\begin{equation}
    \lambda_1 = 1 / (1 + tanh(n))
\end{equation}
\begin{equation}
    \lambda_2 = tanh(n) / (1 + tanh(n))
\end{equation}

After the training is completed, we use the checkpoint with the best performance on the validation set to perform the test on the testing set and obtain the final results.

\section{Results and Discussion}

\begin{table}[htbp]

  \centering
  \caption{Dice coefficients performances on the testing set.G-Grade represents Gleason Grade.}

    \begin{tabular}{cccccccc}
    \hline
    \multirow{2}{*}{Method}&\multirow{2}{*}{GAFL}&\multirow{2}{*}{Re-weighting}&\multirow{2}{*}{Mean dice}&\multirow{2}{*}{Benign}&
    \multicolumn{3}{c}{G-Grade}\cr\cline{6-8}&&&&&{3}&{4}&{5}\cr
    \hline
    PSPNet & & & 55.53 & 90.58  & 54.96 & 76.60 & 0.00\\ \hline
UNet & & & 54.69 & 91.57 & 55.42 & 71.78 & 0.00\\ \hline
DeepLabV3+ & & & 57.56 & 89.81 & \textbf{65.94} & 74.40 & 0.09\\ \hline
HRNet18 & & & \textbf{59.04} & 92.30 & 64.42 & \textbf{78.87} & \textbf{0.58} \\ \hline
HRNet48 & & & 58.51 & \textbf{92.33} & 63.83 & 77.88 & 0.00 \\ \hline \hline
HRNet18 & & yes & 59.40 & 92.78 & 64.40 & \textbf{80.43} & 0.00  \\\hline
HRNet18 & yes & & 63.11 & 92.43 & \textbf{67.15} & 77.63 & 15.22 \\ \hline
HRNet18 & yes & yes & \textbf{66.21} & \textbf{93.26} & 64.49 & 77.74 & \textbf{29.37}\\ \hline \hline
HRNet48 & & yes& 60.31 & \textbf{92.78} & 67.96 & \textbf{80.51}& 0.00\\ \hline
HRNet48& yes& & 61.64 & 91.83 & 64.72 & 73.49 & 16.50 \\ \hline
HRNet48 & yes & yes & \textbf{63.64} & 92.18 & \textbf{68.48} & 72.70 & \textbf{21.19}\\ \hline
    \end{tabular}
    \label{ablation}
\end{table}

We perform a test on several popular backbones, including U-Net~\cite{ronneberger2015u}, DeeplabV3Plus~\cite{chen2018encoder}, HRNet18, and HRNet48~\cite{sun2019deep}. The HRNet18 achieves the best performance, and HRNet48 achieves the second-best. It is worth noting our model also outperforms the PSPNet\cite{zhao2017pyramid}, which won the first place in the Gleason 2019 challenge. This may attribute to more advanced backbone structures. To validate the effectiveness and robustness of the GAFL and label re-weighting modules, we perform ablation studies on both HRNet18 and HRNet48.  The results are summarized in Table.\ref{ablation}. The proposed method learns more information from images with biased labels. Experimental results demonstrate that the two modules consistently improve the performances. It is worth mentioning that since Gleason challenge 2019 is already closed and the performance is judged online by uploading the prediction of an unlabeled dataset. Therefore we are not able to perform fair comparisons with previously reported results.

Compared to the baseline, both the label re-weighting module and GAFL modules can better integrate information from different pathologists to reduce bias from noisy labels. It is interesting to notice that for HRNet, re-weighting only brings slight performance increases (by 0.36\%), whereas GAFL brings a 4.07\% improvement. This may be due to that learning the information of re-weighting is complex and hard. On the other hand, roughness is a well-defined prior knowledge, and it is easy to take effect more stably. It is also worth noting that for both backbones, the GAFL module can help them better focus on the important areas and learn the features of Gleason grade 5. It is interesting to notice that adding both modules can further improve the Dice coefficient by 7.17\% for HRNet18 and 5.13\% for HRNet48. The performance boost is mainly attributed to a dramatic improvement in the segmentation of Gleason grade 5 (from 0.58\% to 29.37\% for HRNet48 and from 0\% to 21.19\% for HRNet48).

\begin{figure}[h]
\centering
\includegraphics[scale=0.45]{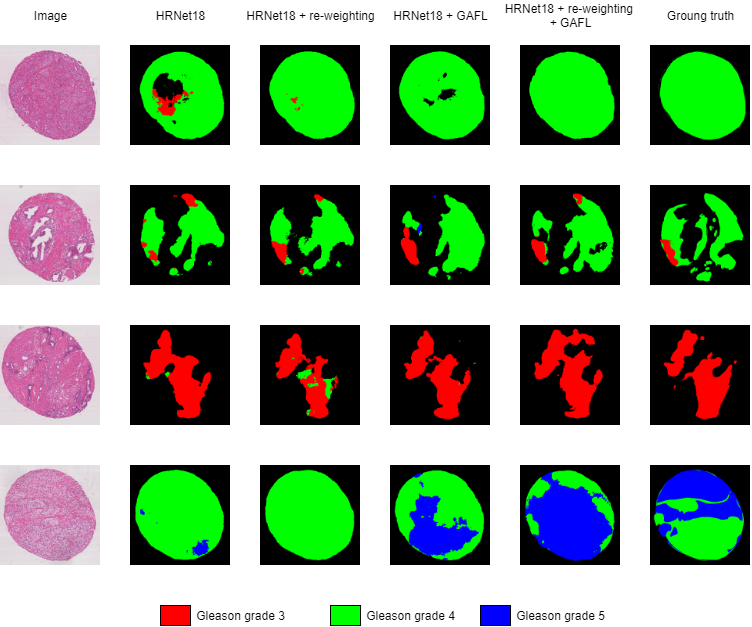}
\caption{Visualization results of the predictions by using different methods. Here red represents Gleason grade 3, green represents Gleason grade 4, and blue represents Gleason grade 5.}
\label{visualization_results}
\end{figure}

We also present some visualization results in Fig.~\ref{visualization_results} to demonstrate the effectiveness of our methods. It shows that both GAFL and re-weighting modules make the segmentation more accurate. 
\section{Conclusion}
This work presents a novel framework to combine multiple experts' annotations for segmentation. A new label re-weighting framework is proposed to account for the reliability of different experts’ labels. A Gaussian attention focal loss is then devised, which takes roughness as prior knowledge to re-weight the focal loss. Our model is tested on the public Gleason 2019 datasets. The results show that our approach effectively improves the model’s robustness against noisy labels regardless of the backbone models being used, and achieves state-of-the-art results. In the future, we plan to further study the theoretic aspect of noisy labels and investigate how to tackle the severely imbalanced sample issue.

\bibliographystyle{splncs04}
\bibliography{citation}
\end{document}